\documentclass[aps,prb,epsf,epsfig,floatfix,showpacs,groupedaddres,superscriptaddress,twocolumn]{revtex4-1}
\bibpunct{[}{]}{,}{n}{}{}
\usepackage{graphicx}
\usepackage[usenames,dvipsnames]{xcolor}
\usepackage{amsfonts}
\usepackage{amsmath}
\usepackage{amssymb}
\usepackage{verbatim}
\usepackage{simplewick}
\usepackage{eucal}
\usepackage{enumerate}
\usepackage{subfigure}
\usepackage{eucal}
\usepackage{url}
\definecolor{uc}{rgb}{0,0.6,0.1}

\usepackage[breaklinks]{hyperref} 
\hypersetup{
     colorlinks=true,
     citecolor=blue,
     urlcolor=orange,
     linkcolor=red
}

\usepackage{mycommand}
\usepackage{natbib}
\usepackage{jrnl}
\def\l{\lambda}
\def\eff{\text{eff}}
\def\ex{\text{ex}}
\def\th{\text{th}}
\def\opt{\text{opt}}
\def\peak{\text{peak}}
\def\HF{\text{HF}}
\def\IPT{\text{IPT}}
\def\CPA{\text{CPA}}
\def\tauinv{\tau^{-1}}
\providecommand{\affiliation}[1]{{\it{\centering #1}}\\}
\providecommand{\email}[1]{\footnote{#1}}
\providecommand{\etal}{{\it et al.}}
\newcommand{\FIGDIR}{.}
\begin{document}
%
%

\author{Himadri Barman}\email{hbarhbar@gmail.com}
\author{Mukul S. Laad}\email{mslaad@imsc.res.in}
\author{Syed R. Hassan}\email{shassan@imsc.res.in}

\affiliation{Institute of Mathematical Sciences, Taramani, Chennai 600113, India
}

\title{Can disorder act as a chemical pressure? An optical study of the Hubbard model}


\date{\today}
%
%
\begin{abstract}
The optical properties have been studied using the dynamical mean-field theory (DMFT) on a disordered Hubbard model. Despite the fact that disorder turns a metal to an insulator in high dimensional correlated materials, we notice that it can enhance certain metallic behavior as if a chemical pressure is applied to the system resulting in an increase of the effective lattice bandwidth (BW). We study optical properties in such a scenario and compare results with experiments where the BW is changed through chemical doping and obtain remarkable similarities vindicating our claim. We also make a point that these similarities differ from some other forms of BW tuned optical effects.  
\end{abstract}

\maketitle

Optical studies have driven a huge attention towards understanding
interaction effects on strongly correlated electronic materials (SCEMs), 
specifically after the discovery of high temperature 
superconductors~\cite{marel:etal:nature03,basov:timusk:rmp05}. 
For the  frequency ($\om$) dependent complex optical conductivity
$\sig\omb=\sig_1\omb + i\sig_2\omb$, real ($\sig_1$) and imaginary ($\sig_2$) parts of it provide much information to probe properties beyond the Drud\'e paradigm of optical conductivity for 
SCEMs. Recently many dynamical quantities related to optical conductivity, particularly the effective carrier density, scattering rate and dynamical effective mass have been found to be useful
in understanding correlated metallic phase in cuprates~\cite{marel:etal:nature03}, pnictides~\cite{moon:etal:prb14,barisic:etal:prb10}, V$_2$O$_3$~\cite{deng:sternbach:haule:basov:kotliar:14}, VO$_2$~\cite{qazilbash:etal:sc07},
organic conductors~\cite{merino:etal:prl08}, ruthenates~\cite{kamal:kim:eom:dodge:prb06,dang:mravlje:georges:millis:prl15,geiger:etal:prb16}, and other correlated materials~\cite{basov:averitt:marel:dressel:haule:rmp11}.

It is known that pressure or doping the transition metals with ions of equal valency but a different size (chemical pressure) leads to a change in the effective bandwidth (BW) of a transitional metal oxide (TMO). In the half-filling carrier concentration such a change may give rise to a Mott metal-to-insulator transition (MIT). Such a transition is often dubbed bandwidth-controlled MIT (BC-MIT)~\cite{imada:fujimori:tokura:rmp98}. Though extensive optical studies have been performed on myriads of TMOs and other SCEMs~\cite{basov:averitt:marel:dressel:haule:rmp11}, studies of disorder-effect on them occupy limited volumes in the literature, both in experiments and theories.
Recently Radonji\'c \etal~\cite{radonjic:etal:prb10} studied a disordered Hubbard model
following the X-ray irradiation induced disorder in $\kappa$-BEDT organic conductor~\cite{sasaki:crystals12}, which is
a typical two-dimensional Mott insulator. However, the authors limited their investigation
to the extended metalicity due to BW increase in presence of disorder, from their transport
and optical results, while how the BW change affects several dynamic properties compared 
to the similar effect in clean systems remains unanswered.
This sets up a motivation to investigate the mentioned optical properties and relate them to BW-controlled physics. 

The site-disordered Hubbard model is written as
\blgn
\h H
=-\sum_{\lngl ij\rngl.\sig} t_{ij}c\y_{i\sig}c\py_{j\sig}
+U\sum_i \nia\nid + \sum_{i\sig}(\eps_c-\mu+v_i) n_{i\sig}\,
\label{eq:ahm:site:disord}
\elgn
where $c\y_{i\sig}$/$c\py_{i\sig}$ is the electron creation/annihilation operator with spin $\sigma$
at site $i$, $t_{ij}$ indicates the amplitude of hopping from site $i$ to $j$ (typically $t_{ij}=t$ $\forall\, i,j$), $U$ is the onsite
Coulomb interaction, $\mu$ and $\eps_c$ are chemical potential and orbital energy of the electrons in clean system, and $v_i$ is the disorder potential at site $i$. The model has been addressed by 
several authors, particularly within the framework of dynamical mean-field theory (DMFT)~\cite{laad:craco:muller-hartmann:prb01,byczuk:hofstetter:vollhardt:prl05,radonjic:etal:prb10,kuleeva:kuchinskii:jetp13,poteryaev:etal:prb15,ekuma:etal:prb15} where
the spectral density shows a disorder driven MIT in a certain parameter regime. 
However, the optical properties comparatively received lesser attention and we investigate 
the dynamic properties studied in the clean system~\cite{merino:etal:prl08} 
in our disordered model.

{\bf\emph{Method} :} We solve \eref{eq:ahm:site:disord} using the DMFT in which a correlated lattice 
model is mapped onto a single impurity Anderson model where the impurity is self-consistently connected to
a non-interacting fermionic bath~\cite{georges:kotliar:krauth:rozenberg:rmp96}. Despite  
the lattice problem gets simplified in this way, the impurity model still requires many-body numerics to be solved and among many such existing methods~\cite{georges:kotliar:krauth:rozenberg:rmp96} we employ the standard iterated perturbation theory (IPT) which is a second order perturbation around the Hartree-Fock (HF) self-energy : 
~\cite{hbar:nsv:ijmpb11} 
\blgn
\Sig_\IPT\omb=\Sig_\HF +\Sig^{(2)}\omb
\label{eq:SE:IPT}
\elgn
with
\blgn
\Sig_\HF&=U\langle \h n_\sigma\rangle=U n/2\,;
\label{eq:SE:HF}\\
\Sig^{(2)}\omb&=\lim_{i\om_n\to\om}\frac{U^2}{\beta^2}\sum_{m,p}\mcG(i\om_n+i\nu_m)\mcG(i\om_p+i\nu_m)\mcG(i\om_p)\,
\label{eq:SE:2}
\elgn
where $n$ is average occupancy, $\beta$ is inverse temperature, $\iom,\iop,\inm$ are relevant Matsubara frequencies, and $\mcG$ is the Weiss Green's function of the non-interacting bath~\cite{georges:kotliar:krauth:rozenberg:rmp96}. In practice, either of $\mcG$ or $\Sig$ is guessed in the first iteration
of the self-consistency loop. The self-consistency condition in DMFT equates both the impurity 
and lattice self-energies and hence the same self-energy in \eref{eq:SE:IPT} is used to calculate the lattice Green's function 
$G\omb=\int d\eps\, D_0(\eps)/(\om+\mu-\eps-\Sig\omb)$
where $D_0$ is the non-interacting lattice density of states and we choose
the Bethe lattice in our work where $D_0(\eps)=2/(\pi t)\sqrt{1-(\eps/t)^2]}$ implying
non-interacting BW in our theory, $\mcW_\th=2t$.
We also work in the half-filling case ($n=1$, $\mu=U/2$) and  
select a binary alloy distribution of the disorder potential 
$P(v_i)=(1-w)\del(v_i)+w\del(v_i-v)$
where $w$ is the weight for disordered sites and for the half-filling case $w=1/2$. The effect of disorder is treated through the coherent potential approximation (CPA)~\cite{velicky:kirkpatrick:ehrenreich:pr68,laad:craco:muller-hartmann:prb01,radonjic:etal:prb10,hbar:laad:hassan:arxiv16}, which gives rise to the following self-energy
and Green's function.
\blgn
    \Sig_\CPA\omb&=wv + \f{w(1-w)v^2}{\om-v(1-w)-\Del\omb}\,,\\
    \mcG_\CPA\omb&=\f{1-w}{\om-\Del\omb}+\f{w}{\om-v-\Del\omb}\,
\elgn
with $\Del\omb$ defined through the Dyson Eq. 
$\Del\omb=\om+\mu-v/2-\Sigma\omb-G\inv\omb$, $\Sig\omb=\Sig_\IPT+\Sig_\CPA$.
$\mcG_\CPA$ becomes the updated Weiss function ($\mcG=\mcG_\CPA$) 
and hence feeds back to \eref{eq:SE:2} in the DMFT self-consistency loop.

To determine the optical conductivity we use the standard expression based on the Kubo formalism :
~\cite{pruschke:jarrell:freericks:ap95,hbar:nsv:ijmpb11}
\blgn
\sig\omb
&=\sig_0 \nint d\om'\, F(\om,\om')\nint d\eps\,\Phi_{xx}(\eps) D_\eps (\om') D_\eps(\om'+\om)\,.
\elgn
where $\sig_0\equiv 4\pi e^2/\hbar$ ($e$ and $\hbar$ being electronic charge and the reduced Planck's constant),  $F(\om,\om')\equiv [f(\om')-f(\om+\om')]/\om$ and $\Phi_{xx}$ 
is called the transport function defined as
\blgn
\Phi_{xx}(\eps)\equiv\f{1}{N}\sk\bigg(\f{d\eps_\kv}{dk_x}\bigg)^2\,\del(\eps-\eps_\kv)\,
\elgn
where $\eps_\kv$ is the momentum ($\kv$) dependent lattice dispersion.
We choose $\Phi_{xx}^{\text{HCL}}(\eps)=D_0(\eps)$ which is exact for the hypercubic lattice at infinite dimension and a reasonable approximation for the Bethe lattice that we consider in our calculations~\cite{lange:kotliar:magnetotr:prl99}. For simplicity, we express $\sig\omb$ in the unit of $\sig_0$, i.e. we set $\sig_0=1$.

\begin{figure}[!htp]
\subfigure[]{
\includegraphics[height=5cm,clip]{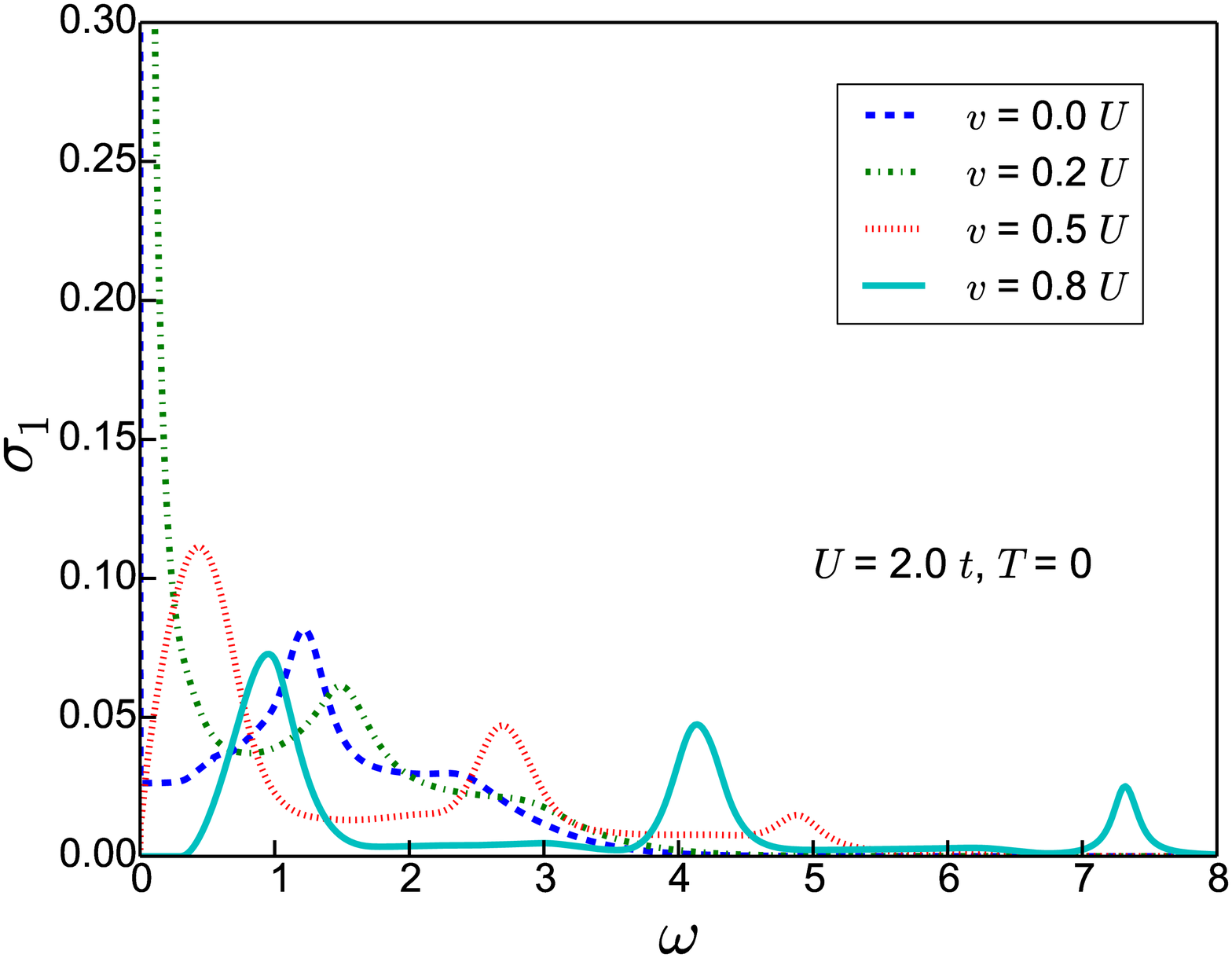}
\label{fig:sig1:T:0}
}
\subfigure[]{
\includegraphics[height=5cm,clip]{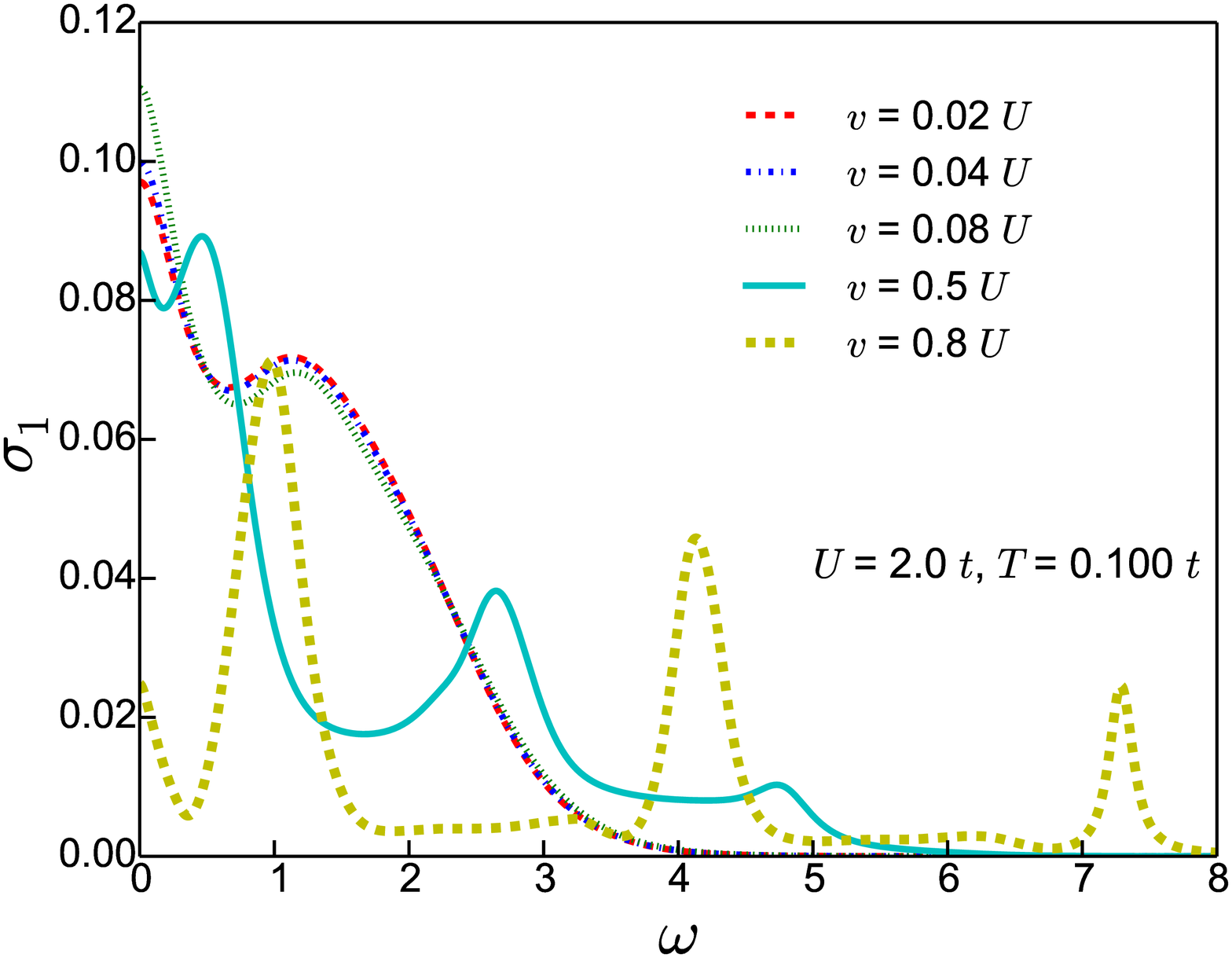}
\label{fig:sig1:T:finite}
}
\caption{Imaginary parts of optical conductivity as functions of frequency ($U=2t$) at (a) zero temperature
and (b) finite temperature. The inset in (a) shows the corresponding spectral densities $A\omb=-\im G\omb/\pi$.} 
\label{fig:sig1}
\end{figure}

{\bf\emph{Results} :} We first look at the real part of the complex optical conductivity : 
$\sig_1\omb\equiv\re\,\sig\omb$. Since disorder induces localization of electrons, and CPA effectively captures such an effect for disordered binary alloys~\cite{laad:craco:muller-hartmann:prb01}, our zero temperature results show that the Drud\'e peak ($\sig_1(0)$) diminishes as disorder strength increases and finally disappears by opening an optical gap at $v>v_c$ ($v_c \simeq 0.5 U$ at $U=2t$, see \fref{fig:sig1:T:0}.), signaling an MIT 
(see inset in \fref{fig:sig1:T:0} for the spectral densities reflecting the same). The Drud\'e peak 
becomes finite at finite temperature ($T$)  and slowly merges with the first absorption
peak as temperature is raised. The optical gap formed at large disorder strength ($v=0.8 U$) gets closed at high temperature ($T=0.1 t$, see \fref{fig:sig1:T:finite}).
%


Though this signifies transition from a good metal to bad metal, 
our interest sticks to the case where disorder leads to the change in the BW 
before it ends up with an Anderson-like insulating phase at $v>v_c$. This bears a close resemblance to the BC-MIT, where higher disorder strength effectively increases the lattice BW and following that it increases the quasiparticle weight $Z=(1-\pd \Sig/\pd \om)|_{\om=0}$ in the coherent regime~\cite{laad:craco:muller-hartmann:prb01,radonjic:etal:prb10,hbar:laad:hassan:arxiv16}.

\begin{figure}[!htp]
\subfigure[]{
\includegraphics[height=5cm,clip]{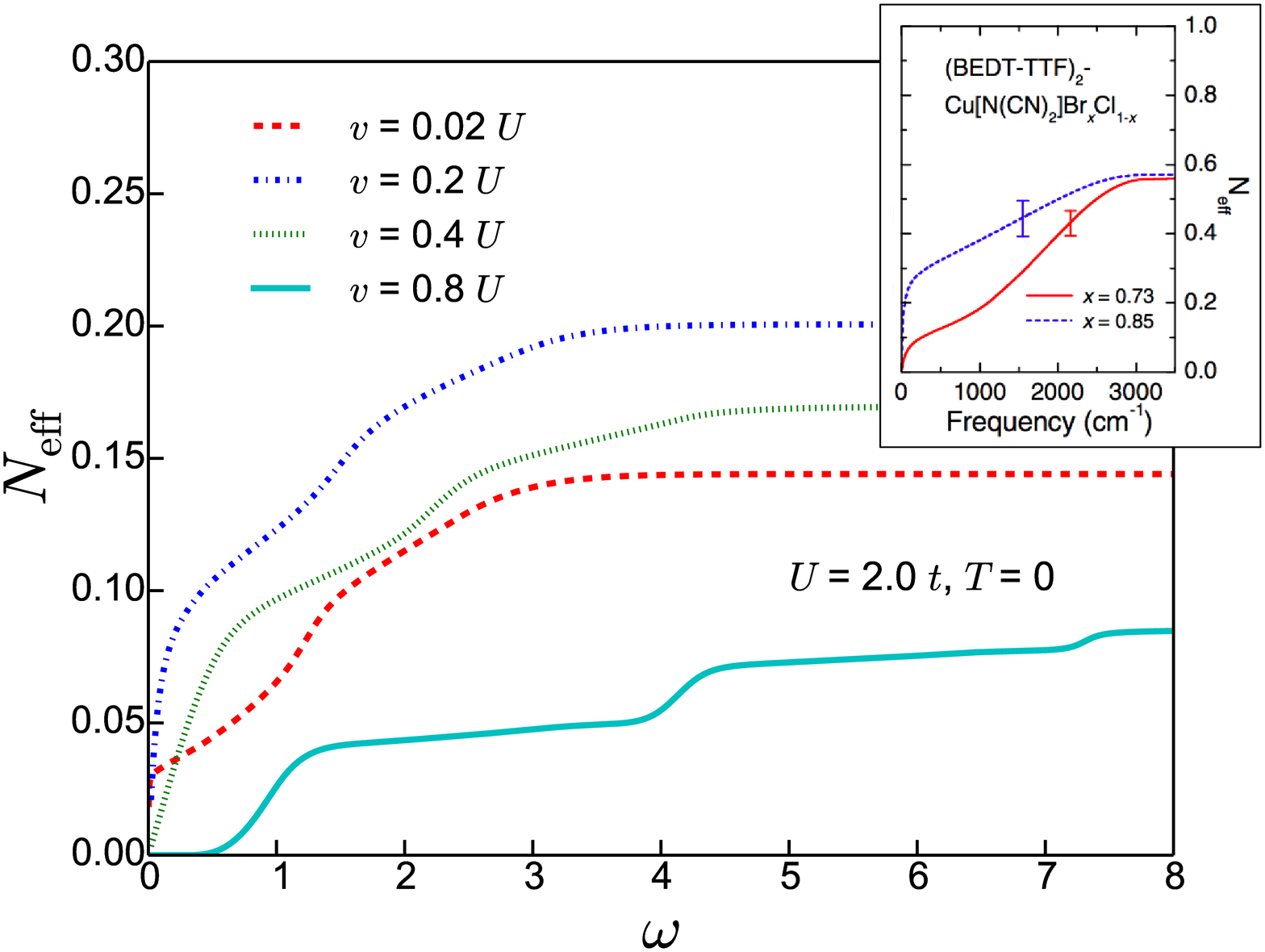}
\label{fig:Neff:T:0}
}
\subfigure[]{
\includegraphics[height=5cm,clip]{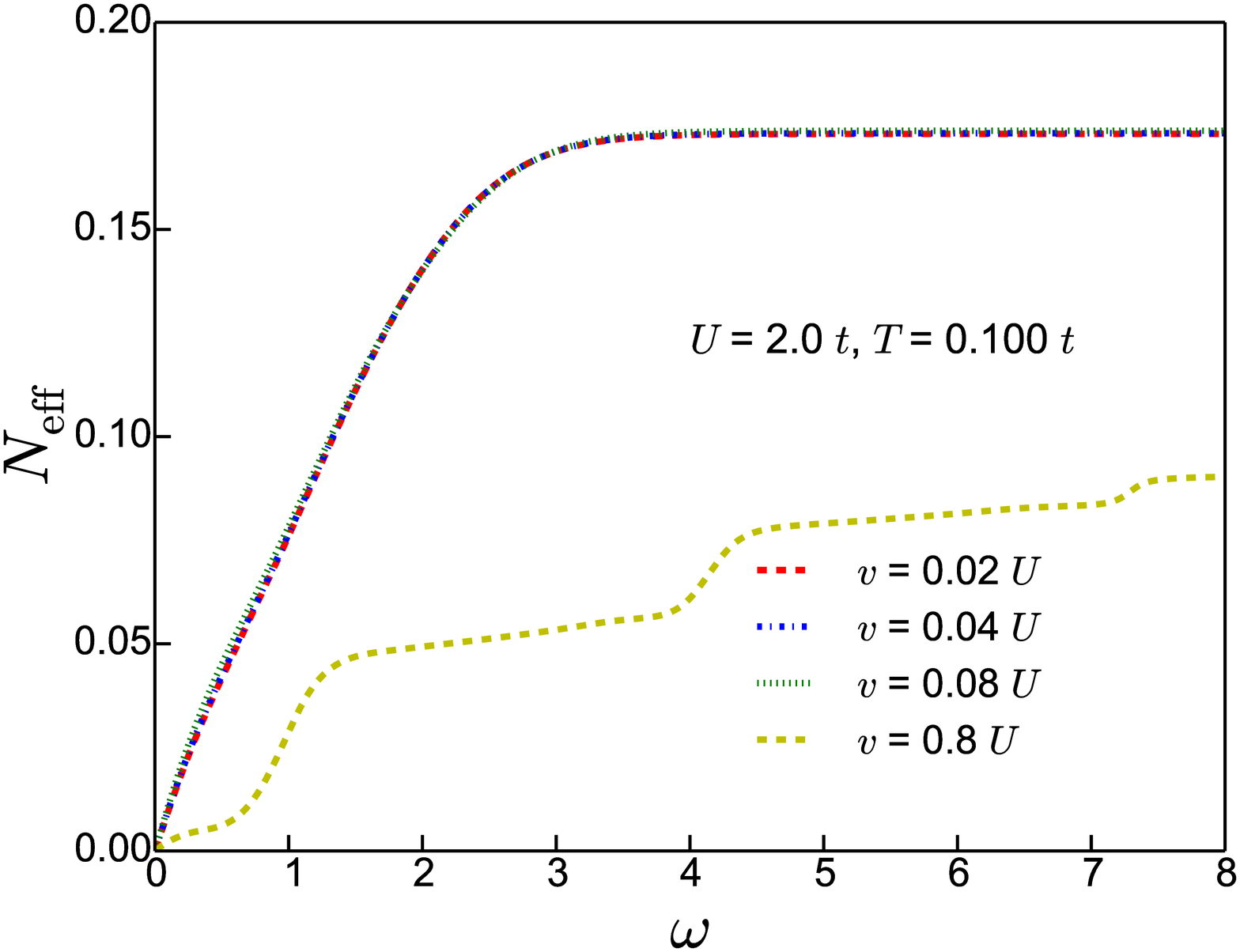}
\label{fig:Neff:T:finite}
}
\caption{(a) Effective carrier density $N_\eff$ as functions of frequency $\om$ at various disorder strengths at $U=2t$. Inset shows experimental results at dopings $x=0.73$ and $x=0.85$ reproduced from ~\citenum{merino:etal:prl08}. Increase in $x$ leads to reduction of $U/t$ i.e. increase in effective BW. (b) $N_\eff$ as functions of frequency $\om$ at $T=0.1 t$.}
\end{figure}

Recently several optical measurements have been performed on the Br-doped organic conductor $\kappa$-BEDT~\cite{merino:etal:prl08} where particularly three kinds of optical property have been investigated, namely (i) effective carrier density ($N_\eff\omb$), (ii) dynamic scattering rate ($\tau\omb$), and (iii) effective optical mass ($m^*\omb$). Following the $f$-sumrule $\zint d\om\, \sig_1\omb=\pi n e^2/(2m)$ for 
the Drud\'e optical conductivity of free electron metals, one can define the effective spectral weight or
charge density $N_\eff$, by performing cumulative sum (instead of full sum upto $\infty$) on $\sig_1\omb$~\cite{basov:averitt:marel:dressel:haule:rmp11} :
\blgn
N_\eff\omb\equiv \f{2m_\opt}{\pi e^2}\int_0^\om d\om'\,\sig_1\ombp\,
\elgn
where $m_\opt$ is called the optical mass of electrons (equivalent to effective mass derived from the bandstructure but in presence of correlation now) for the generic Drud\'e theory~\cite{basov:averitt:marel:dressel:haule:rmp11}.
The other two quantities are defined from the  complex optical conductivity for the generalized Drud\'e model :
%
$4\pi \sig(\om,T)
=\om_p^2/[\tau\inv(\om,T)-im^*(\om,T)/m_\opt]$
($\om_p$ is the plasma frequency), following which 
we find
\blgn
\f{1}{\tau\omb}=\f{\om_p^2}{4\pi}\re\bigg[\f{1}{\sig\omb}\bigg]\,,\label{eq:tau}\\
\f{m^*\omb}{m}=-\f{\om_p^2}{4\pi\om}\im\bigg[\f{1}{\sig\omb}\bigg]\,.\label{eq:mstar}
\elgn
%
%
%
%
\emph{Effective  carrier density -} The main panel in \fref{fig:Neff:T:0} shows how $N_\eff\omb$ changes with $\om$ at various disorder strengths. 
As a generic trend, $N_\eff\omb$ increases with $\om$ as more charge carriers can be excited at higher optical energy. 
%
%
%
%
%
%
%
Increasing $v$ leads to increasing BW and hence the Drud\'e peak gets broadened (see \fref{fig:sig1:T:0})
resulting in more weight in the $f$-sumrule at low frequency. Since for small $v/U$, significant change solely happens around
the Drud\'e peak ($\om/t \lesssim 1$) and hence below a critical value $v_c$, $N_\eff\omb$ is higher 
at higher $v/U$. From \fref{fig:sig1:T:0} one can also speculate that slope change in the optical sum-rule should 
occur at the frequency where there is a significant feature (such as an absorption peak or shoulder) in $\sig_1\omb$. 
Thus the peaks and shoulder at $\om \simeq$ $1.4 t$ and $2.5 t$  manifest change of slopes at the same values for $v=0.02U$, while $U=2t$. At high frequency, since there remains no charge carrier density to be excited: $\sig_1(\om\to\infty)\to 0$, $N_\eff\omb$'s value does not alter much and finally saturates. Similar trends have been noticed in experiments described in Ref.~\citenum{merino:etal:prl08}
(see inset of \fref{fig:Neff:T:0}; $x$ implies doping concentration in the legend).  
%
%
%
%
%
\begin{figure}[!htp]
\subfigure[]{
\includegraphics[height=5cm,clip]{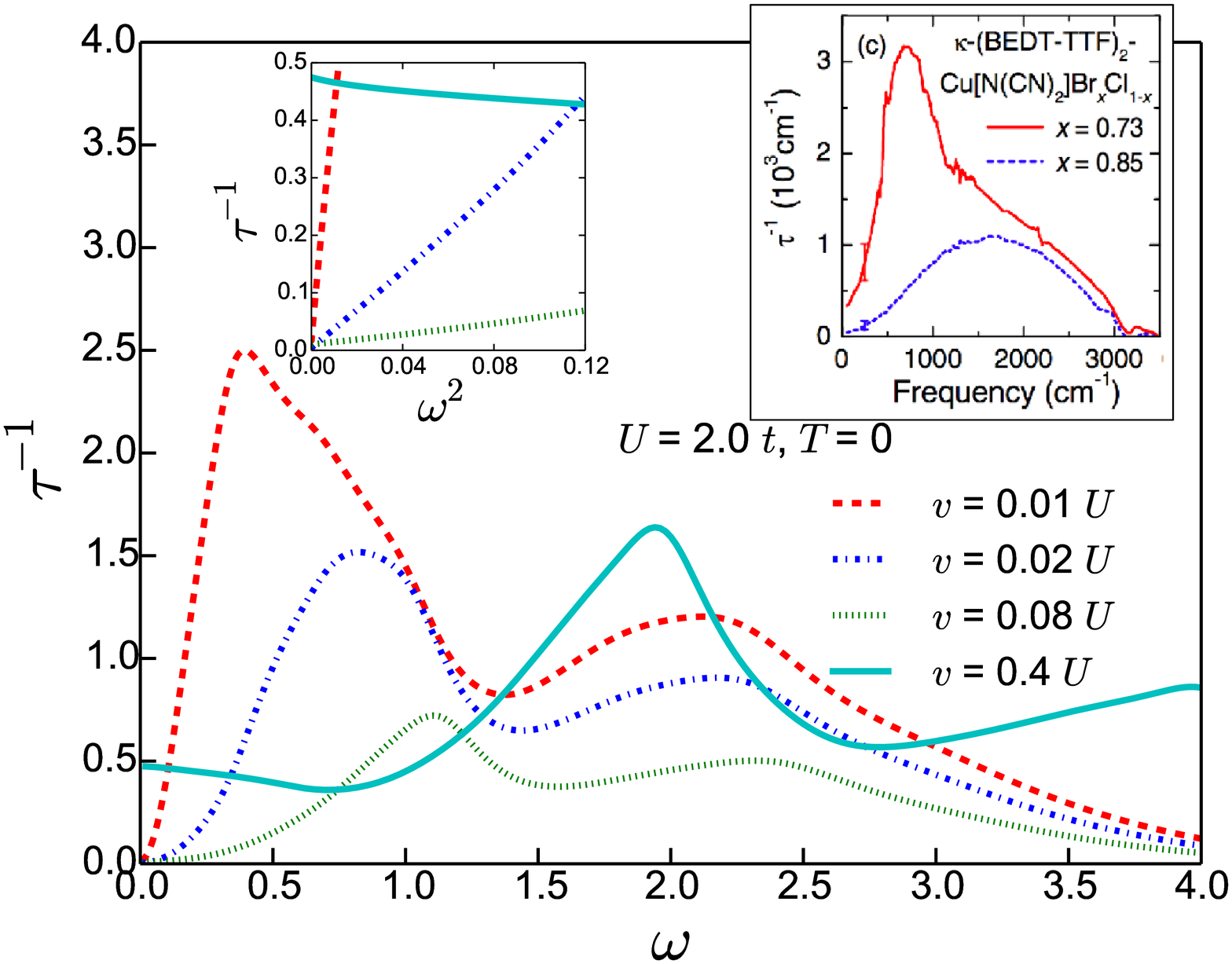}
\label{fig:tauinv:T:0}
}
\subfigure[]{
\includegraphics[height=5cm,clip]{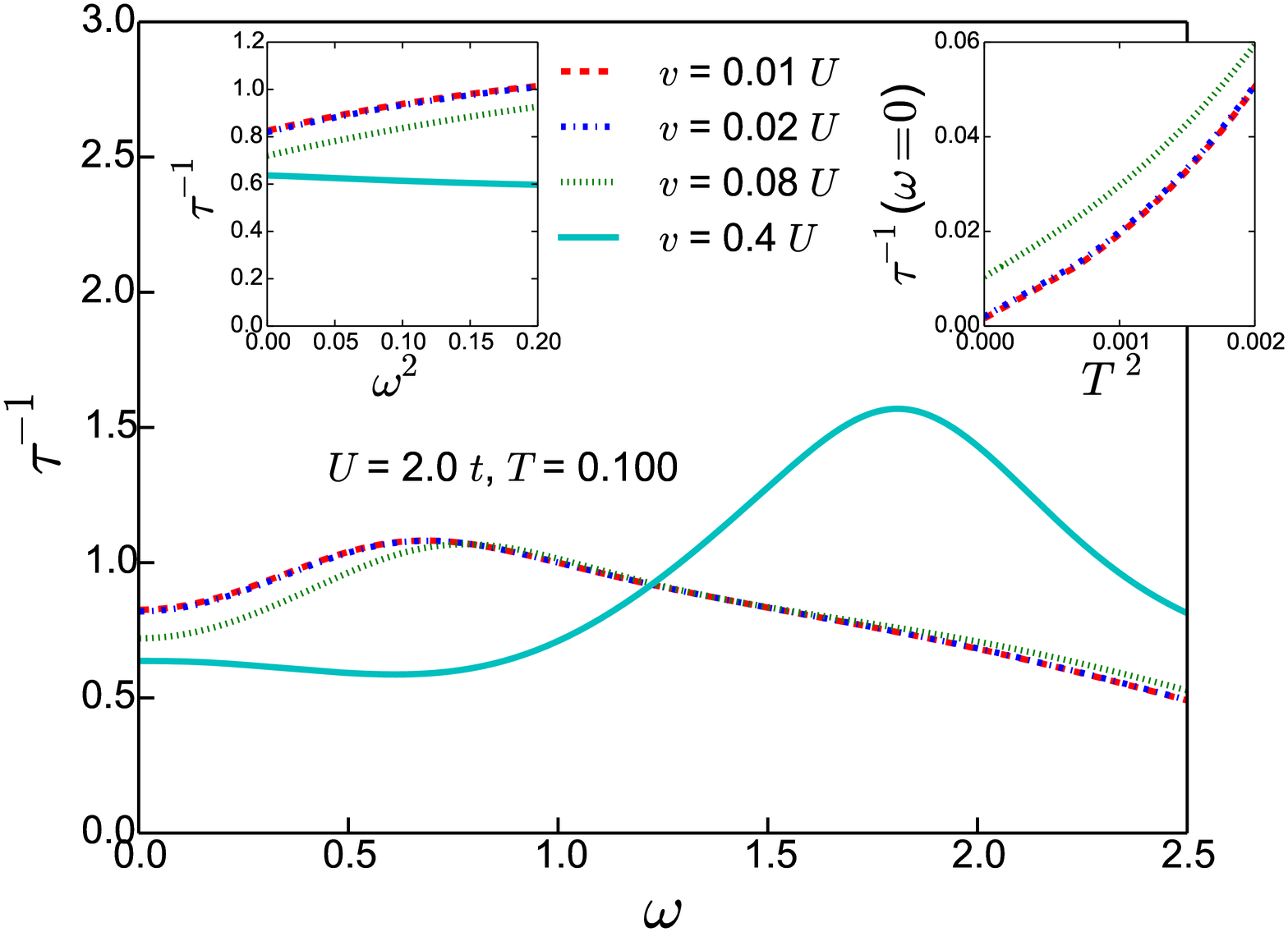}
\label{fig:tauinv:T:finite}
}
\caption{Dynamic scattering rates as functions of frequency ($U=2t$) at (a) $T=0$ and at (b) $T=0.1 t$.
The right inset in (a) shows similarities with the experiment reproduced from Ref.~\citenum{merino:etal:prl08}. The other insets show Fermi liquid frequency 
and temperature dependences and their breakdowns at $v=0.4 U$ and $T>0.01 t$.}
\label{fig:tauinv}
\end{figure}
%
%
%

The feature is more or less the same at finite temperature. However, at quite high temperature ($T> 0.05 t$) the Drud\'e peak significantly melts down and merges with absorption peaks resulting in featureless monotonically increasing $N_\eff$ as a function of $\om$ until it reaches the saturation at large $\om$ (see \fref{fig:Neff:T:finite}). 
In the insulating regime ($v>v_c$), $N_\eff\omb=0$ for $\om < \Del_\opt$ where $\Del_\opt$ is the optical gap. At higher $\om$, $N_\eff\omb$ starts increasing and forms plateaus at positions where the absorption peaks appear in $\sig_1\omb$.   


\emph{Dynamic scattering rate -} 
%
%
%
%
%
%
Next we look at the dynamic scattering rate $\tau\omb$ as defined in \eref{eq:tau}. For a Fermi liquid, it measures the quasiparticle lifetime and it bears the following
frequency-temperature ($\om,T$) dependence.  
\blgn
\tauinv(\om,T)=A\om^2+B(\pi T)^2\,
\label{eq:tauinv:FL}
\elgn
where $A$ and $B$ are constants~\cite{berthod:mravlje:deng:zitko:marel:georges:prb13,stricker:etal:prb14}.
At low or zero temperature, for $v<v_c$, $\tauinv$ depends on $\om$ and $T$ 
in accord with \eref{eq:tauinv:FL} and hence signifies a Fermi-liquid metallic phase. The insets in \fref{fig:tauinv} show the plots against $\om\sq$ and $T\sq$ supporting
the fact. For $v\ge v_c\simeq 0.4 U$, $\om\sq$ dependence and $T>0.01 t$, $T\sq$ dependence get violated, insinuating breakdown of FL regime. 
At zero temperature a mid-infrared peak cum shoulder feature arises in the scattering rate while the shoulder is not much evident in the experiment reported in Ref.~\citenum{merino:etal:prl08}. However, the DMFT results in the same reference (BW tuned by $U/t$ ratio) contain the shoulder feature reinforcing the fact BW tuning by both $U/t$ and $v/U$ are of the same ilk. 
The peak position appears around 750 $\cminv$ and 1500 $\cminv$ respectively
for Br-doping $x=0.73$ and $x=0.85$ respectively, which are $0.30 \mcW_\ex$ and $0.62 \mcW_\ex$
considering the experimental noninteracting BW, $\mcW_\ex=0.3$ eV $\simeq 2419.66\, \cminv$~\cite{merino:etal:prl08}. Similarly, though disorder is a different drive
compared to the experiment, the peaks for $v=0.01 U$ and $v=0.08 U$ at $U=2t$ occurs 
around  $0.5t=0.25 \mcW_\th$ and $1.2 t = 0.6 \mcW_\th$ respectively, which are remarkably within the same energy range.
Blueshifting of the peak due to increase in BW also agrees with the experiment. At sufficiently high temperature ($T=0.1t$) the shoulder feature disappears followed by a long universal tail extending to high frequency.
After $v\ge 0.4 U$, the quasiparticle description entirely breaks 
down and $\tauinv$ starts decreasing as frequency increases (see \fref{fig:tauinv}).
%

%

%
\emph{Effective optical mass -}
%
Optical mass is another interesting property that tells about the renormalization
of the electronic mass as a combined effect of electronic bandstructure and correlation. Like in Ref.~\citenum{merino:etal:prl08} the dynamical mass ratio $m^*\omb/m$ at low $\om$ also decreases due to increase of the BW by rise in disorder strength. While the real part contribution $\tau\inv\omb$ of the conductivity in  \eref{eq:tau} gives rise to a peak feature, the  contribution from its imaginary part (\eref{eq:mstar}) forms a dip in the mid-infrared frequency range. 
In the experiment described in Ref.~\citenum{merino:etal:prl08} the dip
occurs around 1000 $\cminv$ ($=0.413 \mcW_\ex$) 1900 $\cminv$ ($=0.785 \mcW_\ex$)
for dopings ($x$) 0.73 and 0.85 respectively. Similarly for $v=0.01U$ and $v=0.08U$,
 the dips appear at $0.8t=0.4 \mcW_\th$ and $1.4 t = 0.7 \mcW_\th$ respectively, which
are again in the same energy range. As the BW increases further due to increase in $v/U$, the dip experiences a blueshift and becomes shallower. 
At $v\ge 0.4 U$ the dip continues blueshifting, however, the ratio 
increases with frequency at low $\om$ regime instead of decreasing, again signaling a breakdown of Fermi liquid coherence.
%
%
%
%
%
%
%
\begin{figure}[!htp]
\subfigure[]{
\includegraphics[height=5cm,clip]{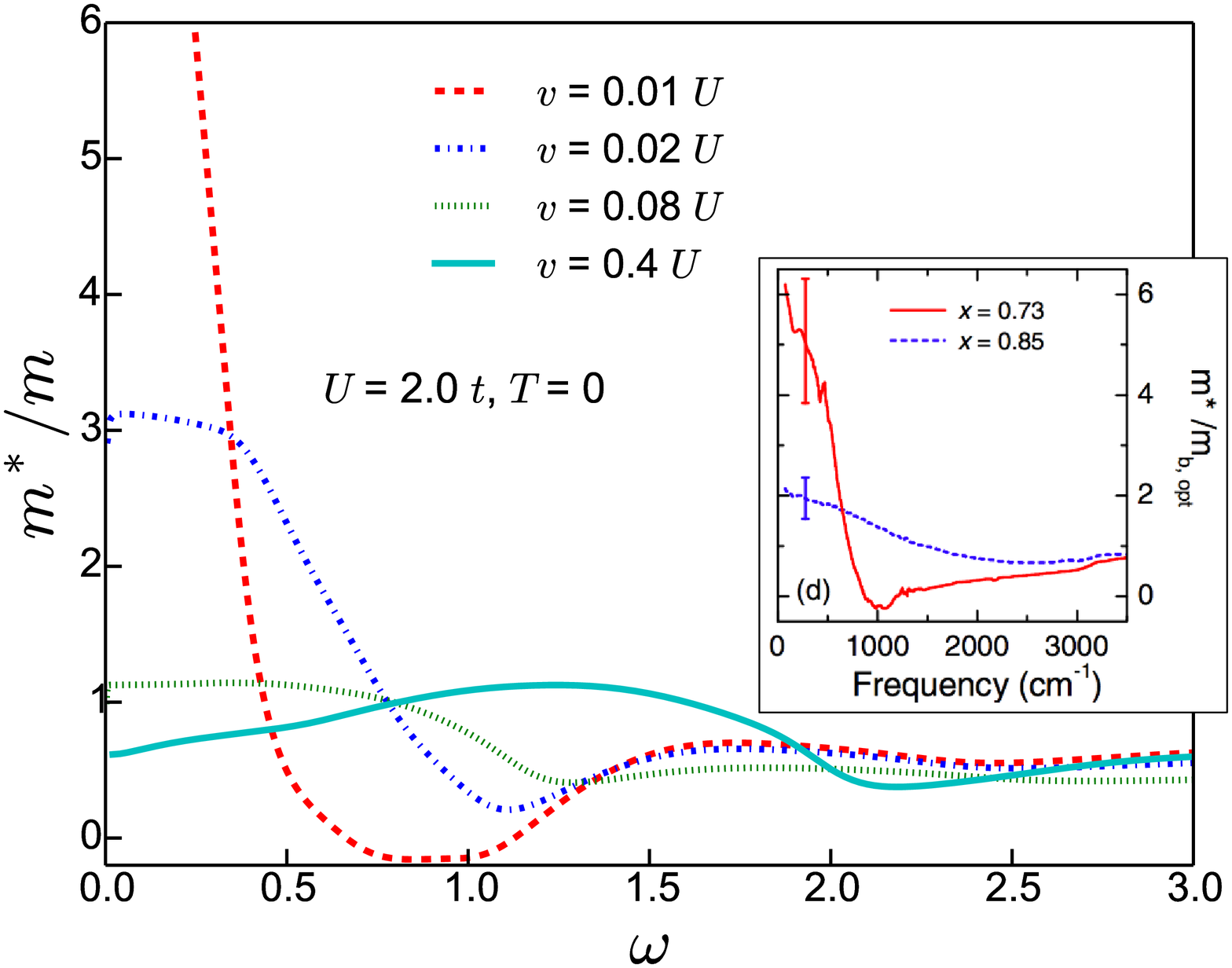}
\label{fig:dynmass:T:0}
}
\subfigure[]{
\includegraphics[height=5cm,clip]{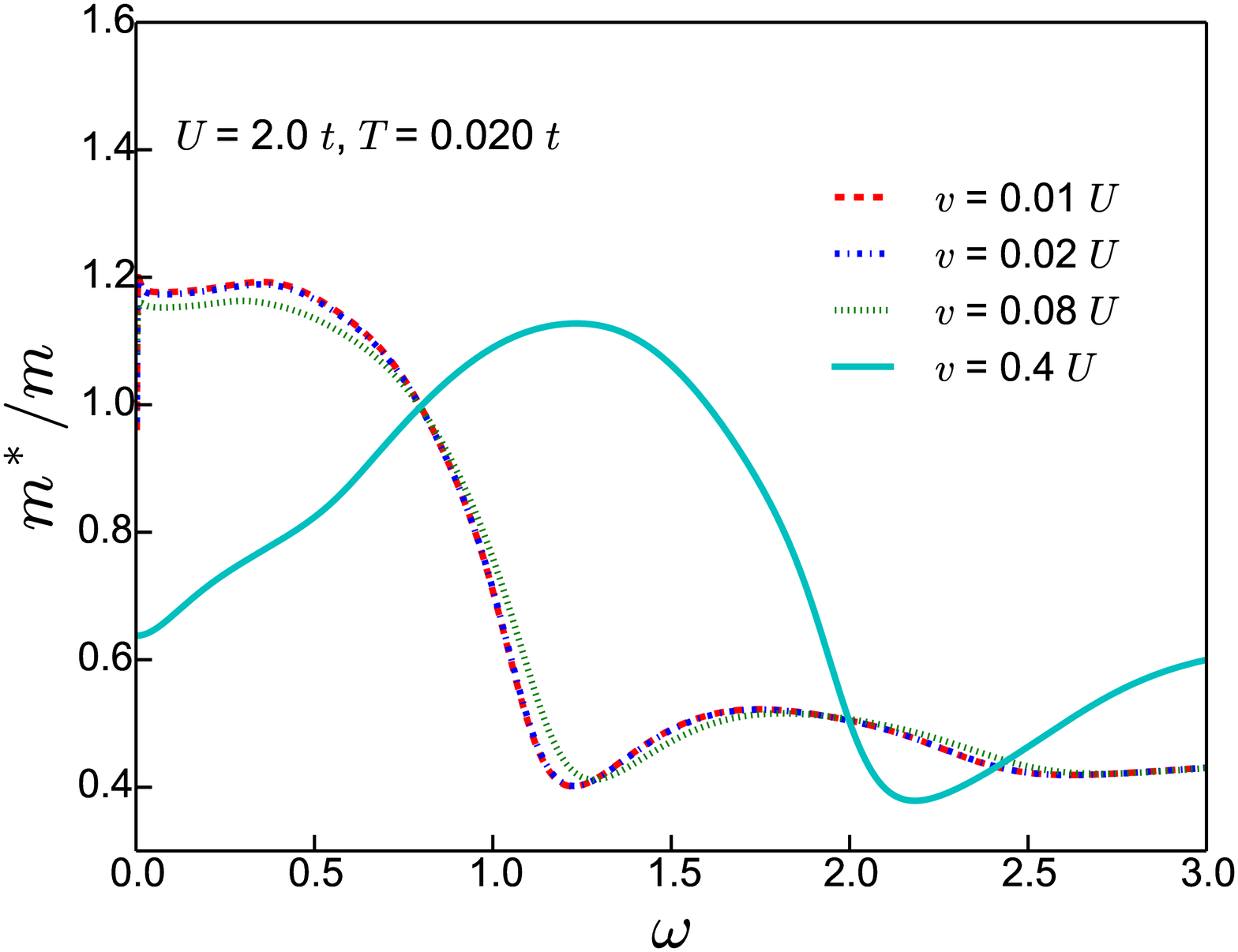}
\label{fig:dynmass:T:finite}
}
\caption{Dynamical effective masses as functions of frequency ($U=2t$.) at (a)
zero temperature and (b) finite temperature. At $v=0.4 U$, the Fermi-liquid property
breaks down. The right inset in (a) shows similar experimental results reproduced 
from Ref.~\citenum{merino:etal:prl08}.} 
\label{fig:dynmass}
\end{figure}
%
%
%
%
%
%
%
%
At $T\ge 0.02 t$, $m^*\omb/m$ value drops down near $\om=0$ and
almost collapses for all frequency range for various $v$'s less than $0.4 U$ (see \fref{fig:dynmass:T:finite}). 

\emph{Other BW controlling factors - } There exist several factors which can control the effective BW of a correlated lattice and for fixed $t$, we mention four such important ones, viz.  (i) Coulomb interaction $U$, (ii) carrier doping $\del=1-n$, (iii) temperature $T$ and (iv) disorder potential $v$.   
%
%
%
%
%
%
%
%
%
%
%
\begin{figure}[!htp]
\subfigure[]{
\includegraphics[height=5cm,clip]{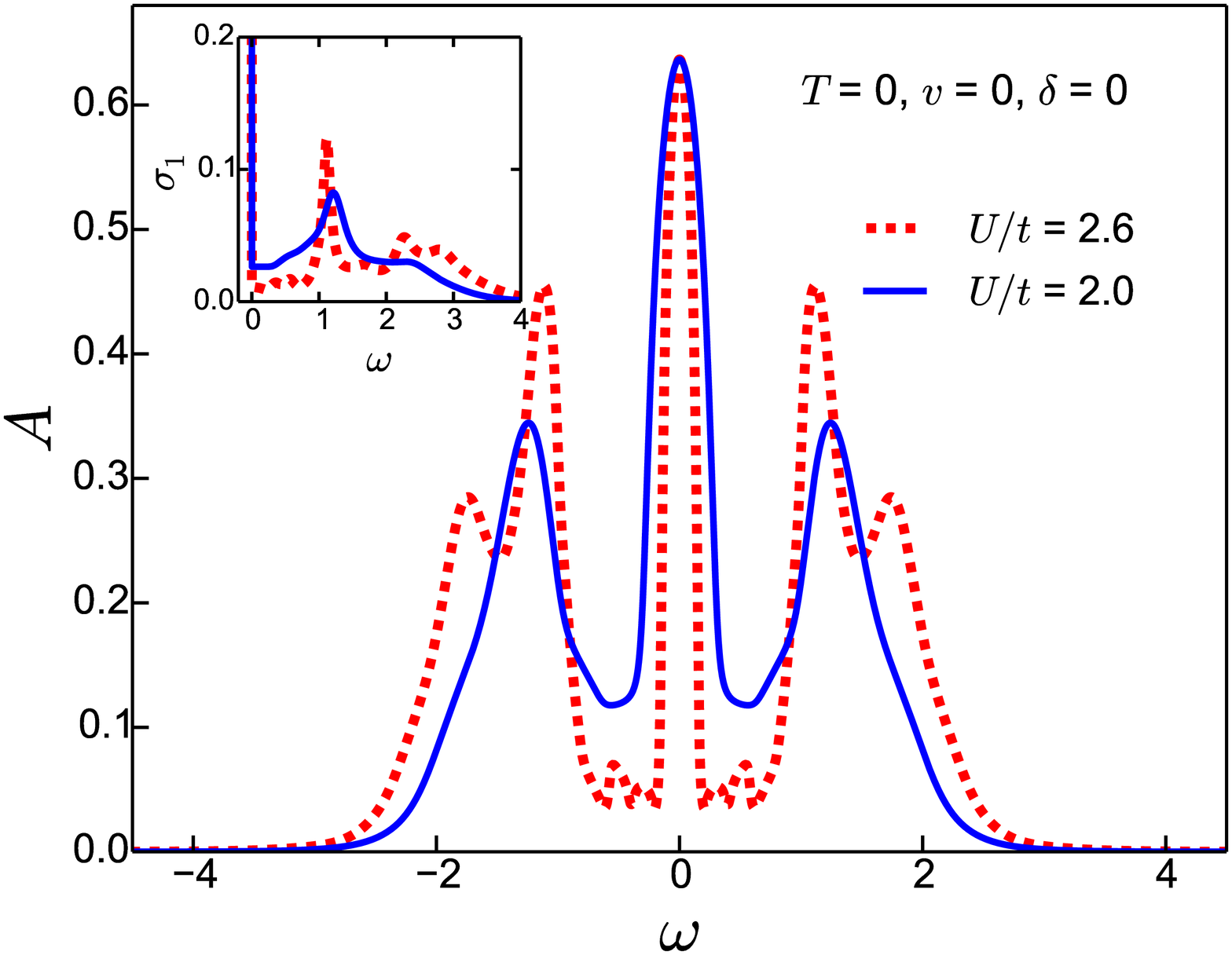}
\label{fig:Uvary}
}
\subfigure[]{
\includegraphics[height=5cm,clip]{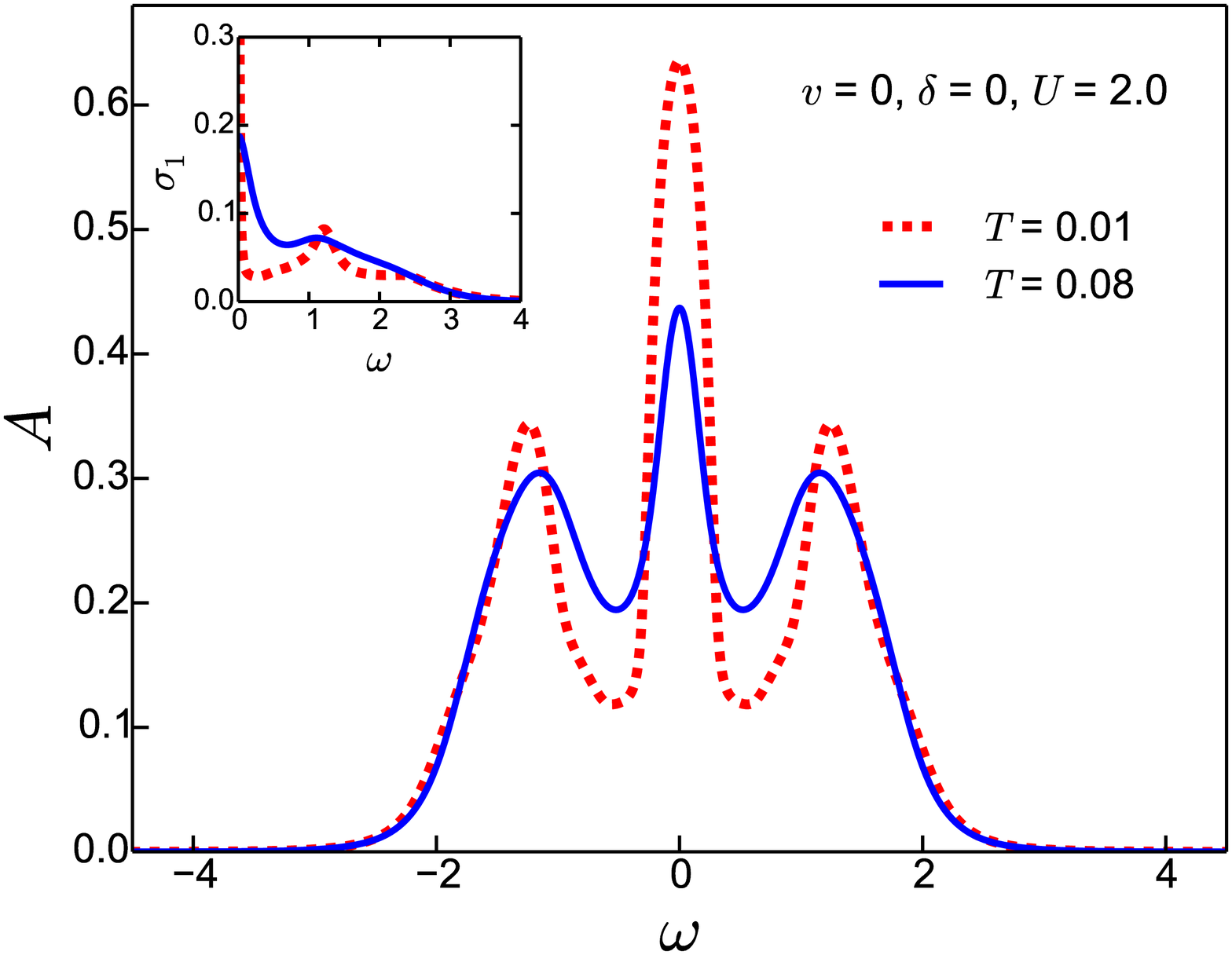}
\label{fig:Tvary}
}
\subfigure[]{
\includegraphics[height=5cm,clip]{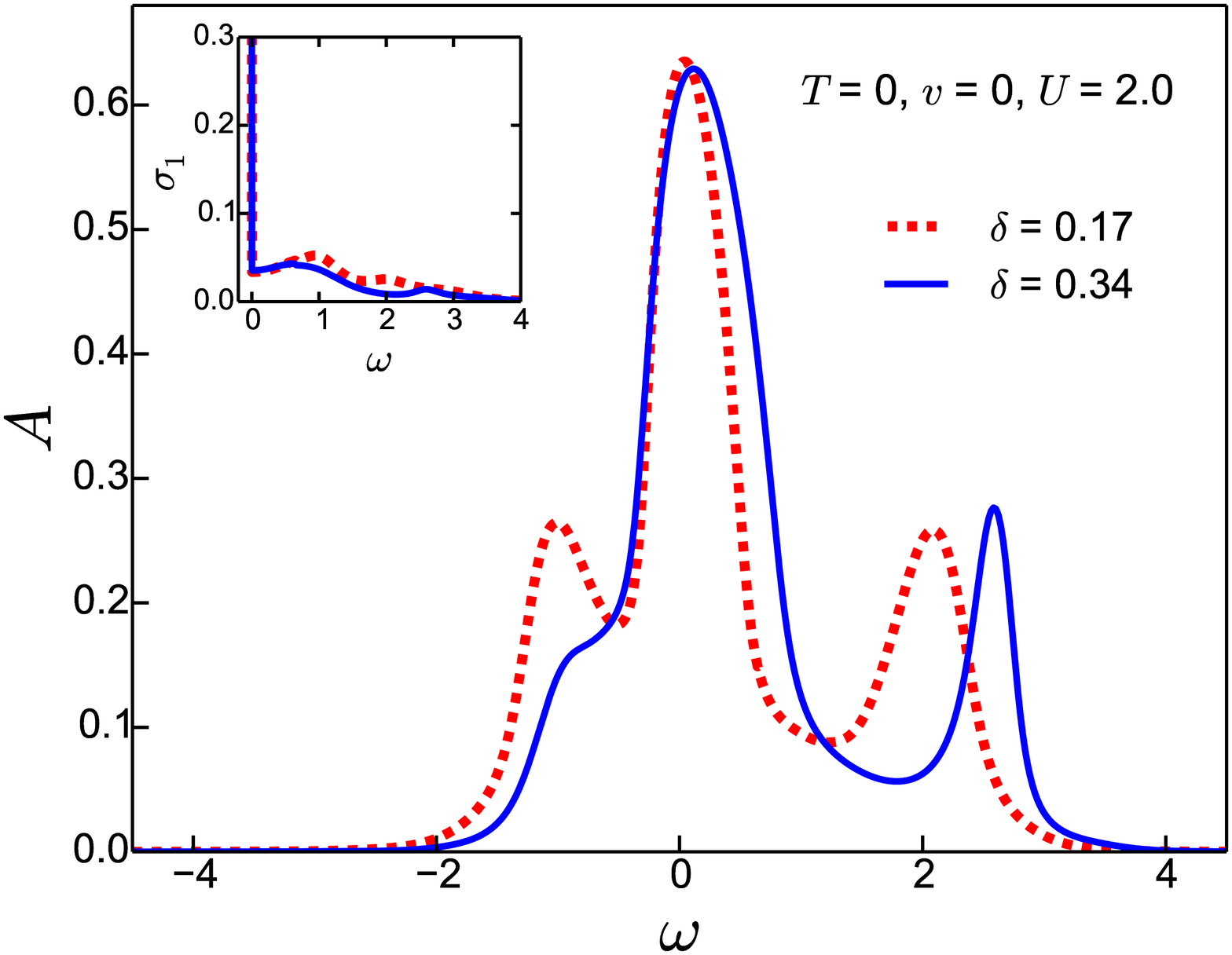}
\label{fig:delvary}
}
\subfigure[]{
\includegraphics[height=5cm,clip]{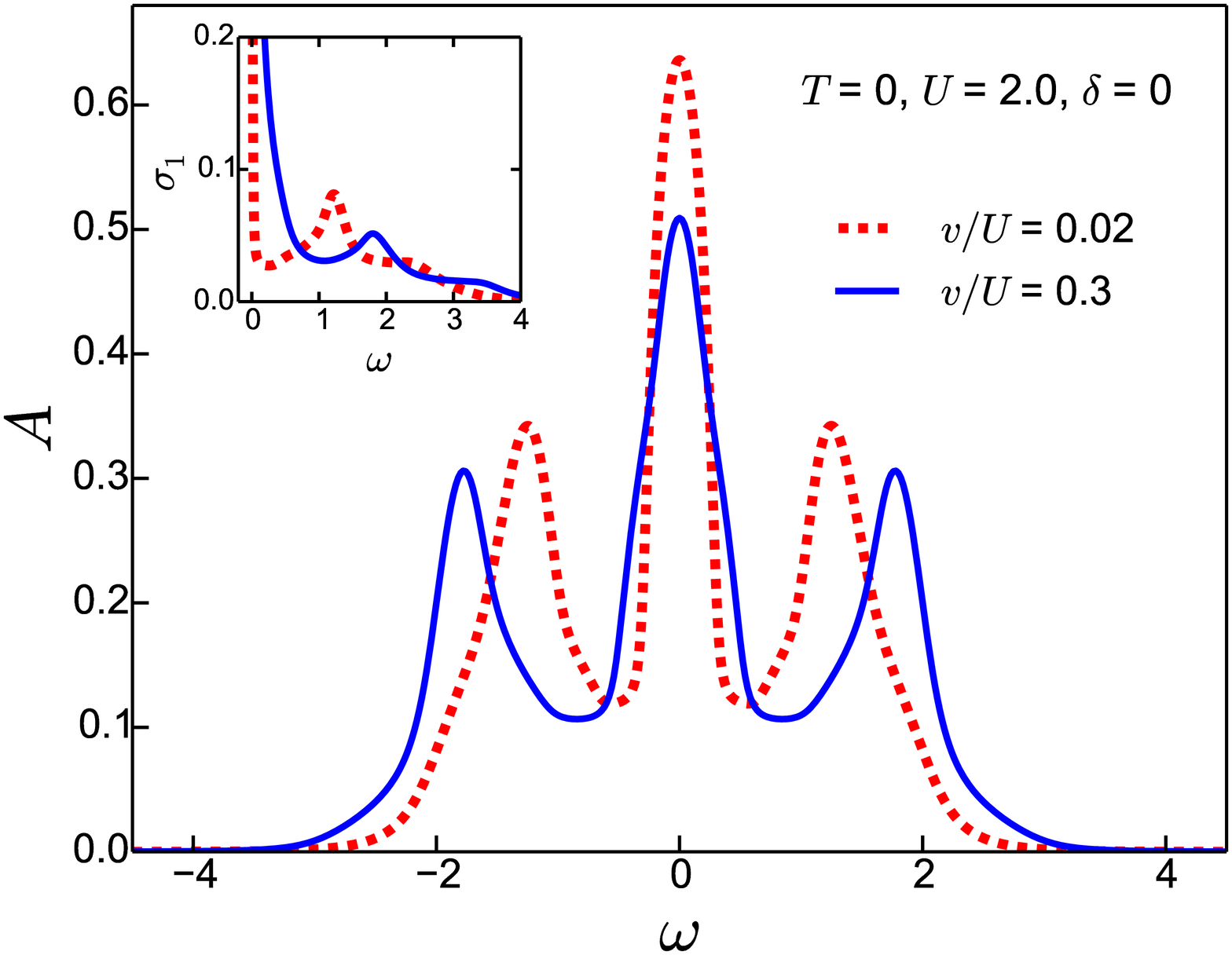}
\label{fig:vvary}
}
\caption{Modification of spectral bandwidth due to (a) decrease in $U/t$,
(b) increase in $T$, and (c) increase in doping $\del=1-n$, and increase in disorder potential $v$. 
The insets show
corresponding changes in optical property $\sig_1\omb$.}
\label{fig:other:factors}
\end{figure}
%
%
%
%
%
At fixed $t$, decreasing $U$ lowers 
the ratio $t/U$ and effectively increases BW.  
The width of the quasiparticle peak in the spectral density $A\omb$
represents a measurement of the effective BW. \fref{fig:Uvary}   
shows the width enhances as $U$ is reduced from $2.6 t$ to $2.0 t$, however,
keeping the height at the Fermi level ($\om=0$) 
unchanged due to the \emph{Luttinger pinning}~\cite{vollhardt:etal:jpsj05} property 
of a Fermi liquid. The pinning remains intact as well in the 
Drud\'e weight of optical conductivity ($\sig_1(\om \to 0)$) 
though the mid-infrared peak position $\om_\peak$ acquires a blueshift and broadening as
$U/t$ is reduced. 
Similarly increasing $T$ also broadens the quasiparticle peak though 
the Luttinger pinning does not hold any more.  
As a consequence of this, the Drud\'e peak in $\sig_1\omb$ also diminishes. 
However, $\om_\peak$ redshifts as the BW increases due to temperature
rise.
Deviation from the particle-hole symmetry again leads to change in BW. For small $\del$,
the Luttinger pinning is obeyed and effective BW increases as $\del$ increases. Like in the 
$T$-driven case, BW increasing leads to redshift in $\om_\peak$. 
Now if we look back to the calculations with changing disorder, we can see the absorption peak
goes to a blueshift like the $U/t$-driven BW increase. This ensures that disorder indeed 
acts like chemical pressure from the change in BW perspective.
The comparisons can be viewed in \fref{fig:other:factors},
which are summarized in the table below.\\

\begin{tabular}{|l | c | c |}
\hline
Factor &Effect on BW & Effect on $\om_\peak$\\
\hline
\hline
Pressure ($t/U$) & Increase & Blueshift\\
\hline
Disorder ($v/U$) & Increase & Blueshift\\
\hline
Temperature ($T$) & Increase & Redshift\\
\hline
Doping ($\delta$) & Increase & Redshift\\
\hline
\end{tabular}

\smallskip
{\bf\emph{Summary} :}
In this work, we for the first time establish the fact 
that the interaction/pressure and disorder driven bandwidth (BW) changes play very similar roles on optical properties, while 
other alternatives such as carrier filling and temperature change lead to different behaviors even though 
both shape the BW. Our results are at par with the 
the experiment on Br-doped BEDT conductor~\cite{merino:etal:prl08}. 
Though the experiment has been practically done in clean samples,
our results invites similar experiments on disordered correlated systems~\cite{sasaki:crystals12}.
Our investigation also could be generalized for a generic disorder distribution and to invoke 
the effect of Anderson localization the same DMFT formalism could be combined with the typical
medium theory (TMT)~\cite{dobrosavljevic:pastor:nikolic:epl03}. On the DMFT side, thought IPT provides a reliable insight, recently developed exact impurity solvers such as 
continuous time Monte Carlo (CTQMC) method could be implemented 
to find more accurate results and compare to our predictions~\cite{gull:millis:lichtenstein:rubtsov:troyer:philipp:rmp11}.

\section*{Acknowledgments}
HB owes to the DAE, Govt. of India for providing financial support and scientific resources. For the energy unit conversions we used the table provided by NIST, USA.

\newpage
\bibliographystyle{apsrev4-1}
\bibliography{refs}

\end{document}